# Phonon Spectrum in Hydroxyapatite: Calculations and EPR Study at Low Temperatures

Timur Biktagirov, Marat Gafurov, Kamila Iskhakova,
Georgy Mamin, Sergei Orlinskii

*Kazan Federal University, 420008 Kazan, Russian Federation*

**Abstract** Density functional theory based calculations within the framework of the plane-wave pseudopotential approach are carried out to define the phonon spectrum of hydroxyapatite $Ca_{10}(PO_4)_6(OH)_2$ (HAp). It allows to describe the temperature dependence of the electronic spin-lattice relaxation time $T_{1e}$ of the radiation-induced stable radical $NO_3^{2-}$ in Hap, which was measured in X-band (9 GHz, magnetic field strength of 0.34 T) in the temperature range T = (10-300) K. It is shown that the temperature behavior of $T_{1e}$ at T > 20 K can be fitted via two phonon Raman type processes with the Debye temperature $\Theta_D \approx 280$ K evaluated from the phonon spectrum.

## 1 Introduction

One of the major characteristics of the materials in the solid phase is the type and parameters of its phonon spectrum. The phonon spectrum is necessary to analyze and calculate many physical properties of solids - optical, thermal, electrical, etc. Though it is common to approximate a phonon spectrum by parabolic density of states, the study of complex solid compounds could require the knowledge of the true vibrational properties.

Apatites (in particular hydroxyapatites, HAp) were always considered as a very challenging class of material for analytical characterization. It is known that crystallinity, thermal stability, solubility and other physicochemical and biological properties of HAp are strongly influenced by both its morphology and the content of impurities [1]. A new wave of interest to the apatite based materials due to their potential and realized applications especially in biomedical area forces to try some new tools and approaches for their comprehensive analysis. They include semi-empirical and first-principles calculation methods to extract the phonon related quantities [2-4]. Due to the complexity of the chemistry of HAp, there remain apparent discrepancies and contradictions between calculations and experimental results (see [5], for example). Verifications and

T. Biktagirov, M. Gafurov, K. Iskhakova

refinements of the exploiting models should be done and new approaches for the extracting the macroscopic parameters from the computational models should be introduced.

The present work aims to make a connection between first-principles calculations of phonon properties of hydroxyapatite and electron paramagnetic resonance (EPR) spectroscopy. It is known that the relaxation characteristics of radicals in the host crystal are sensitive to the electron spin – phonon interactions and can be used to define vibrational properties of crystals [6]. Various types of carbonate, phosphorous, hydroxyl and oxygen radicals that exist inherently, introduced artificially or generated by external irradiation are fruitfully exploited for the investigations of synthetic and biological apatites by EPR spectroscopy [7-10]. Here we use the results of density functional theory (DFT) based calculations of the phonon spectrum to analyze the temperature dependence of the longitudinal electronic relaxation time $T_{1e}$ of the radiation-induced $NO_3^{2-}$ radical. This spin probe was initially introduced in a trace amount (less than $10^{18}$ spins per gram) as $NO_3^-$ anion during the wet synthesis procedure.

**2 Materials and Methods**

DFT calculations of phonon properties were carried out within the framework of plane-wave pseudopotential approach as implemented in the Quantum ESPRESSO code [11]. The Perdew-Burke-Ernzerhof functional (PBE) was employed [12] together with Vanderbilt ultrasoft pseudopotentials [13] and a cutoff energy of 40 Ry. The results were obtained for 1×1×1 monoclinic supercell of 88 atoms (space group $P2_1/b$). During geometry optimization the cell dimensions and atomic positions were allowed to be fully relaxed with the convergence condition on forces of $10^{-3}$ Ry/Bohr. The Brillouin Zone (BZ) integration was performed on a Monkhorst-Pack 2×2×1 k-point mesh [14]. The phonon density of states (DOS) was calculated on a 2×2×2 q-point grid.

The $NO_3^-$ species in the trace amounts were incorporated into the structure of hydroxyapatite powder with the chemical formula $Ca_{10}(PO_4)_6(OH)_2$ to obtain the crystalites with the average sizes of 1 μm according to equation

$10Ca(NO_3)_2 + 6(NH_4)_2HPO_4 + 8NH_4OH =$
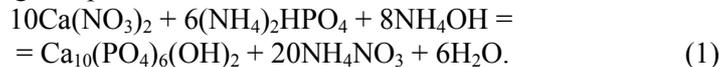
$= Ca_{10}(PO_4)_6(OH)_2 + 20NH_4NO_3 + 6H_2O.$ (1)

The incorporation was done during the wet preparation procedure in the Group of V.I. Putlyaev (Department of the Material Sciences, Moscow State University). Details of the synthesis, post-synthesis treatments and analytical characterization of HAp powders are given in our previous papers [15-19].

Phonon Spectrum in Hydroxyapatite

Pulsed and cw EPR measurements were done using X-band (9 GHz) Bruker Elexsys 580 spectrometer equipped with the liquid helium temperature controller. Electron spin echo (ESE) EPR spectra were recorded by means of field-swept two-pulse echo sequence $\pi/2 - tau - \pi$ by sweeping the external magnetic field $B_0$ with the pulse length of $\pi$ pulse of 16 ns and time delay $tau = 240$ ns. To extract the spin-lattice (longitudinal) relaxation times $T_1$ the Inversion-Recovery pulse sequence $\pi - T_{delay} - \pi/2 - tau - \pi$ with the $T_{delay}$ varied, was applied.

**3 Results and Discussion**

The influence of lattice vibrations on the relaxation rate is conventionally described in terms of Debye model [6] taking into account only acoustic phonons. It allows to fit the experimental EPR results adequately even for the high-temperature superconducting cuprates [20, 21]. This simple approximation implies a parabolic phonon spectrum confined by the Debye frequency $\omega_D$ (or corresponding Debye temperature $\Theta_D$) which is the one of the important parameters in solid-state physics. The *ab initio* calculated phonon DOS of HAp presented in Fig. 1 is definitive not Debye-like. Our DFT results resemble those obtained by Slepko and Demkov for 6×4×6 Monkhorst-Pack k-point mesh for BZ in the VASP code [3].

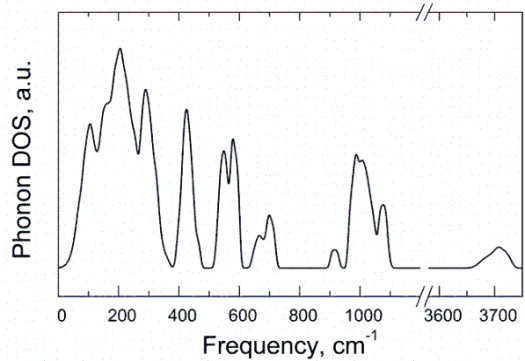

**Fig. 1** *Ab initio* calculated phonon DOS for hydroxyapatite.

Following Anderson [22] and Bjerg [23] we consider the Debye temperature extracted from the whole first-principles calculated phonon spectrum by assigning only three (acoustic-like) vibrational degrees of freedom:

$$\Theta_D = n^{-1/3} \sqrt{\frac{5\hbar^2}{3k_B^2} \frac{\int_0^\infty \omega^2 g(\omega) d\omega}{\int_0^\infty g(\omega) d\omega}} \quad , \tag{2}$$

where $n = 88$ is the number of atoms per vibrational unit (i.e. per unit cell), $k_B$ is the Boltzmann constant, $\hbar$ is the reduced Plank constant, and $g(\omega)$ is phonon density with $\omega$ in the units of angular frequency. It gives $\Theta_{D(calc)} = 280$ K. This value is in remarkable agreement with the experimental $T_{1e}$ data for the carbonate radical $CO_2^-$ in HAp [24]. At the same time, it is much lower than the numbers extracted from the elastic constants and thermal properties measurements (about 543 K and 390 K, correspondingly, see [25]). To check the applicability of this approach at least for the understanding of the EPR results we have conducted our own relaxation measurements on other type of EPR spin-probe as described above.

Fig. 2 shows the ESE detected EPR spectrum of the $NO_3^{2-}$ radiation induced radicals due to the hyperfine interaction between the stable nitrogen-containing paramagnetic center with electronic spin with $S = \frac{1}{2}$ and one nitrogen nuclear spin with $I = 1$ for $^{14}N$ isotope. From the previous EPR data analysis it was concluded that $NO_3^{2-}$ radiation induced radical substitutes $PO_4^{3-}$ position in the HAp structure while the carbonate ions could also replace OH group [15-18].

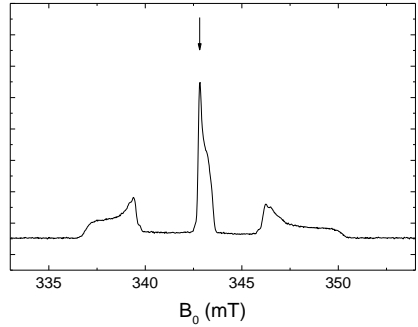

**Fig. 2** Field-swept ESE EPR spectrum of $NO_3^{2-}$ radiation induced radicals in HAp powder at T = 50 K. Arrow marks the value of the magnetic field $B_0$ for which the relaxation times were measured.

The $T_{1e}$ measurements were done in the magnetic field $B_0 = 343$ mT (see Fig. 2). The corresponding data are presented in Fig. 3 along with the approximation according to Eq. (3) with $\Theta_D = 280$ K:

Phonon Spectrum in Hydroxyapatite

$$T_{1e}^{-1} = T_{dir}^{-1} + T_{Raman}^{-1} = C_{dir}T + C_{Raman}T^9 f(\frac{\Theta_D}{T}), \qquad (3)$$

where the first term is responsible for the direct process of relaxation at low temperatures, the second one defines the Raman spin-lattice relaxation process with Debye approximation for lattice vibrations, $C_{dir}$ and $C_{Raman}$ are coefficients which values were fitted, $f(\frac{\Theta_D}{T}) \equiv f(z) \equiv I_8(z)/I_8(\infty)$,

$$I_8(z) = \int_0^z x^8 e^x / \left(e^x - 1\right)^2 dx.$$

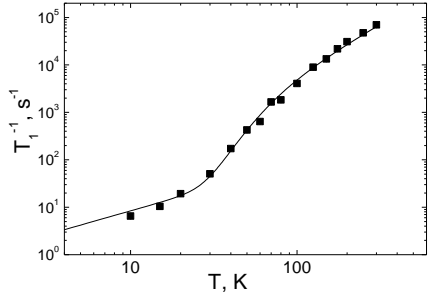

**Fig. 3** Results of the $T_{1e}$ measurements. Solid line is a function of Eq. (3) with $C_{dir}$ = 1.8 s$^{-1}$K$^{-1}$, $C_{Raman}$ = 4.4·10$^{-9}$ (s$^{-1}$K$^{-9}$) and $\Theta_D$ = 280 K.

As seen from the excellent coincidence of the experimental and fitted data, the value of $\Theta_D$ defined from the whole phonon spectrum according to Eq. (2) can adequately describe the EPR derived Raman relaxation.

**4 Conclusion**

For the first time the temperature dependence of spin-lattice relaxation time $T_{1e}$ of the radiation induced nitrate radicals $NO_3^{2-}$ was measured. It is shown that the temperature behavior of $T_{1e}$ at T > 20 K is governed by two phonon Raman type processes. Even though it is shown that HAp is not a Debye-type solid, these processes can be described within the Debye model with the Debye temperature $\Theta_D \approx 280$ K extracted from the *ab initio* calculated phonon spectrum.

**Acknowledgements**
We dedicate this investigation to our teacher, colleague, and friend Dr. Igor N. Kurkin who has been actively engaged in EPR research at Kazan University since the beginning of 1960s and who is going to celebrate his 75$^{th}$ birthday in 2016 in front of an EPR spectrometer. This work is supported by the subsidy allocated to Kazan Federal University for the project part in the sphere of scientific activities. M.G. thanks the support of the Program of competitive Growth of Kazan Federal University among World's Leading Academic Centers.